
\input phyzzx
\def\Mij{M_{ij}}
\def\sigij{\sigma_{ij}}

\Pubnum={CERN-TH-7040/93\cr
hepth@xxx/9310095}
\date={October 1993}
\titlepage
\medskip
\title{Exact Spectrum of $SU(n)$ Spin Chain with Inverse--Square Exchange}
\bigskip
\author {
Alexios P. Polychronakos\footnote\dagger
{poly@dxcern.cern.ch}}
\address{Theory Division, CERN\break
CH-1211, Geneva 23, Switzerland}
\bigskip
\abstract{
The spectrum and partition function of a model consisting of $SU(n)$
spins positioned at the equilibrium positions of a classical Calogero
model and interacting through inverse--square exchange are derived.
The energy levels are equidistant and have a high degree of degeneracy,
with several $SU(n)$ multiplets belonging to the same energy eigenspace.
The partition function takes the form of a $q$--deformed polynomial.
This leads to a description of the system in terms of an effective
parafermionic hamiltonian, and to a classification of the states in terms
of ``modules" consisting of base--$n$ strings of integers.
}
\bigskip
\line{CERN-TH-7040/93 \hfill}
\line{October 1993 \hfill}
\vfill
\endpage

\def\PL{{\it Phys. Lett.\ }}

\def\PRB{{\it Phys. Rev. B\ }}
\def\PRA{{\it Phys. Rev. A\ }}
\def\PRL{{\it Phys. Rev. Lett.\ }}

\def\JMP{{\it J. Math. Phys.\ }}
\def\ADM{{\it Adv. Math.\ }}

\def\JSP{{\it J. Stat. Phys. \ }}

\def\JPA{{\it J. Phys. A.\ }}

\def\LNC{{\it Lett. Nuovo Cimento \ }}
\def\PRep{{\it Phys. Rep.\ }}

\REF\Cal{F.~Calogero, \JMP {\bf 10} 2191 (1969); {\bf 10} 2197 (1969);
{\bf 12} 419 (1971); \LNC {\bf 13}, 411 (1975); F.~Calogero and C.~Marchioro,
\LNC {\bf 13}, 383 (1975).}
\REF\Sut{B.~Sutherland, \PRA {\bf 4}, 2019 (1971); {\bf5}, 1372 (1972);
\PRL {\bf34}, 1083 (1975).}
\REF\Mos{J.~Moser, \ADM {\bf 16}, 1 (1975).}
\REF\OlPe{M.A.~Olshanetsky and A.M.~Perelomov, \PRep {\bf 71}, 314 (1981);
{\bf 94}, 6 (1983).}
\REF\Hal{F.D.M.~Haldane, \PRL {\bf 60}, 635 (1988); {\bf 66}, 1529 (1991).}
\REF\Sha{B.S.~Shastry, \PRL {\bf 60}, 639 (1988); {\bf 69}, 164 (1992).}
\REF\Ino{V.I.~Inozemtsev, \JSP {\bf 59}, 1143 (1989).}
\REF\Haetal{F.D.M.~Haldane, Z.N.C.~Ha, J.C.~Talstra, D.~Bernard and
V.~Pasquier, \PRL {\bf 69}, 2021 (1992).}
\REF\FoMi{M.~Fowler and J.A.~Minahan, \PRL {\bf 70}, 2325 (1993).}
\REF\Pol{A.P.~Polychronakos, \PRL {\bf 70}, 2329 (1993).}
\REF\Fra{H.~Frahm, \JPA {\bf 26}, 473 (1993).}
\REF\HaHa{Z.N.C.~Ha and F.D.M.~Haldane, \PRB {\bf 46}, 9359 (1992).}
\REF\Kaw{N.~Kawakami, \PRB {\bf 46}, 1005 (1992); {\bf 46}, 3191 (1992).}
\REF\MiPo{J.A.~Minahan and A.~P.~Polychronakos, \PL {\bf B 302}, 265 (1993).}
\REF\HiWa{K.~Hikami and M.~Wadati, \PL {\bf A 173}, 263 (1993).}
\REF\SuSh{B.~Sutherland and B.~S.~Shastry, \PRL {\bf 71}, 5 (1993).}
\REF\Ext{All the relevant formulae and relations used in this context
can be found in H.~Exton, ``$q$-Hypergeometric Functions and Applications,"
Ellis-Horwood/Wiley, 1983.}

\chapter{Introduction}

Systems with interactions of the inverse-square type seem to enjoy a
revived popularity. The archetype of such systems is the family of the
Calogero-Sutherland-Moser integrable systems of particles in one dimension
[\Cal-\Mos].
(For an extensive review and a comprehensive list of earlier references
see [\OlPe].) A discrete variant of such systems are spin chain models with
inverse-square exchange. The original such system is the Haldane-Shastry
model [\Hal,\Sha], consisting of $SU(2)$ spins on an equispaced lattice on
the circle, interacting through two-body exchange terms inversely
proportional to the square of the chord distance. Algebraic aspects of this
model have been studied [\Ino,\Haetal], and its integrability was shown
[\FoMi]. Recently, a new model of this type was introduced [\Pol] in which
the lattice points lie at the equilibrium positions of classical Calogero
particles on a line. Some partial and numerical results on this system were
presented in [\Fra]. Furthermore, composite models of this type have appeared
[\HaHa-\HiWa], consisting of generalizations of the Calogero or Sutherland
models for particles with spin.

One of the main points of [\Pol] was the observation that the Haldane-Shastry
model can be thought of as the high-interaction limit of the Sutherland
system with internal degrees of freedom. The spatial degrees of freedom
in this limit decouple and the remaining spin degrees of freedom constitute
the desired spin chain lattice. This fact was exploited in [\SuSh] in order
to solve the Haldane-Shastry model in the thermodynamic limit (number of sites
$\to \infty$). In this paper, we will use it to fully solve the spin chain
model proposed in [\Pol].

\chapter{Review of the Calogero system of particles with $SU(n)$ spins}

We briefly review here the Calogero-type system of particles with spin
and its spectrum, which was first introduced and solved in [\MiPo].
Consider $N$ particles on the line with internal
degrees of freedom (``spin") transforming under the fundamental of $SU(n)$.
The hamiltonian of the system is
$$
H=\half\sum_i ( p_i^2 +\omega^2 x_i^2 )+\sum_{i<j}
{l(l-\Mij)\over(x_i-x_j)^2},
\eqn\ham$$
where $\Mij$ is the operator that exchanges the positions of particles $i$ and
$j$. Define also the operators $\sigij$ which exchange the spins of particles
$i$ and $j$. In terms of the fundamental $SU(n)$ generators for each particle,
$\sigma_i^a$ ($a$=1,2,...$n^2-1$), the $\sigij$ have the expression
$$ \sigij = {1 \over n} +\sum_a \sigma_i^a \sigma_j^a ~.
\eqn\sig$$

For the above hamiltonian, there are raising and lowering operators
$a_i^\dagger$ and $a_i$, defined
$$
a_i^\dagger = p_i +\sum_{j\neq i}{i\over x_i - x_j}\Mij +i\omega x_i ~~,~~~~
a_i = p_i +\sum_{j\neq i}{i\over x_i - x_j}\Mij -i\omega x_i ~,
\eqn\aa$$
satisfying
$$
[a_i , a_j ] = [a_i^\dagger = a_j^\dagger ] = 0 ~~,~~~~
[a_i , a_j^\dagger ] = 2\omega \delta_{ij} \Bigl( 1+l\sum_{k\neq j} M_{ik}
\Bigr) -2\omega (1-\delta_{ij} ) M_{ij} ~,
\eqn\aaa$$
$$
[H,a_i]=-\omega a_i ~~,~~~~ [H,a_i^\dagger] = \omega a_i^\dagger ~.
\eqn\Haa$$
Therefore, the energy eigenstates of $H$ can in principle be found in a
systematic way once the ground state is determined. Since $H$ is invariant
under total particle permutation, we can choose the states to be bosonic.
On such states the total exchange operator $\Mij\sigij$ becomes one, and
thus $\Mij=\sigij$ and the above hamiltonian becomes
$$
H=\half\sum_i ( p_i^2 +\omega x_i^2 )+\sum_{i<j}{l(l-\sigij)
\over(x_i-x_j)^2} ~.
\eqn\hamm$$
This hamiltonian commutes with the total spin $S^a = \sum_i \sigma_i^a$.
We shall call the case $l>0$ ferromagnetic and the one $l<0$ antiferromagnetic.

(Note that in [\MiPo] the choice of fermionic states was made. Due to the
singular nature of the interaction in \hamm, there is no particle penetration
and thus the results will not depend on this choice; the penetrability of the
original hamiltonian \ham, which was due to the exchange operators $\Mij$,
has been traded for a flow between different spin sectors.)

In the ferromagnetic regime, the ground state is annihilated by all $a_i$
and acquires the form
$$
\psi_f = \prod_{i<j} | x_i - x_j |^l \exp\left(-\half\sum_i \omega
x_i^2 \right) \chi (S) ,
\eqn\gsf$$
where $\chi$ is any component of the fully symmetric representation
in the tensor product of fundamentals $\sigma_1 \sigma_2 \cdots \sigma_N$,
that is,
the $N$--fold symmetric irrep of $S$. Thus, the ferromagnetic ground state
$\psi_f$ is $(n+N-1)! \over (n-1)! N!$--fold degenerate.
Excited states can be obtained by acting on the ground states with
symmetric combinations of the raising operators, that is, polynomials
of the operators
$$
A_k \equiv \sum_i (a_i^\dagger )^k ~~~~{\rm and}~~~~
A_k^a \equiv \sum_i s_i^a (a_i^\dagger )^k ,
\eqn\AA$$
which satisfy the commutation relations
$$
[A_k , A_r ] = [A_k , A_r^a ] = 0 ~~,~~~~
[A_k^a , A_r^b ] = f^{abc} A_{k+r}^c ~,
\eqn\AAA$$
with $f^{abc}$ the structure constants of $SU(n)$.
Both $A_k$ and $A_k^a$ create $k$ quanta of energy, but $A_k^a$ also alters
the $SU(n)$ representation by moving one box in its Young tableau.
For an explicit treatment of the $SU(2)$ case, see [\MiPo].

The problem of determining the energy eigenstates can be substantially
simplified by the following observation: the above procedure is,
in fact, identical to the one in the free particle case (without couplings).
Indeed, the ground state \gsf\ is in the same representation as the free
one (take $l=0$), and the free excited states can be found by acting with the
same creation operators $A_k$ and $A_k^a$ (note that the commutation
relations \AAA\ of these operators are independent of $l$). In the free case,
the internal degrees of freedom simply become dynamically irrelevant particle
``flavors." Thus, the states of the system can be found by considering
$N$ free bosons in an external harmonic oscillator potential, each one
being in one of $n$ possible flavors. If $N_k$ such bosons are in the
same level of the harmonic oscillator spectrum, they obviously transform
under the $N_k$--fold symmetric irrep of $SU(n)$. The representation of the
full state then is the tensor product of the symmetric representations
for each filled level. (In the ground state all particles are lumped together
in the lowest level, thus $N_k =N$ and we get the irrep of \gsf.) The only
difference with the free system is an overall shift of the spectrum $E_o$,
accounting for the different ground state energy.

In the case $l<0$ the above ground state \gsf\ is unacceptable due to the
negative power in the exponent of $|x_i - x_j|$. Notice, however, that the
antiferromagnetic hamiltonian \hamm\ can be achieved by turning $l$ into
$-l>0$ and choosing the statistics of the particles to be fermionic
($\Mij \sigij = -1$). Therefore, generalizing the result of [\MiPo] for the
$SU(2)$ case (see also [\HaHa]), the antiferromagnetic ground state is
$$
\psi_{af} = \prod_{i<j} |x_i - x_j |^{-l} \psi_F (\{x_i ,s_i^a \}) ,
\eqn\gsaf$$
where $\psi_F$ is the ground state of a system of $N$ free fermions with
$n$ flavors in an external harmonic oscillator potential.
This state is in the $m$--fold antisymmetric representation of $SU(n)$,
where $m = N\, {\rm mod}\, n$. The excited states can be found by acting
with the same totally symmetric operators as in the ferromagnetic case.
Again, we can explicitly find these states
by considering the states of the corresponding free fermion problem and
shifting all energy levels by the corresponding ground state energy. In
each oscillator level we can put $N_k$ fermions, where $N_k$ is at most $n$.
Such a multiplet transforms now under the $N_k$-fold antisymmetric irrep of
$SU(n)$, and the total state carries the tensor product of irreps for all
occupied oscillator levels. The ground state is $n! \over (n-m)! m!$--fold
degenerate.

An interesting observation is that the states \gsf\ and \gsaf\ are actually
valid only for $|l|>\half$ (this is related to the ``spurious" states in
[\MiPo]). The reason is that hermiticity of the hamiltonian
requires the particle current to vanish at coincidence points $x_i = x_j$,
and this will happen if the power of $|x_i - x_j |$ is greater than $\half$.
In fact, since the eigenstates of the ferromagnetic and antiferromagnetic
cases are quite different, an intricate level-crossing must take place
in the region $|l|<\half$ in order to match the two. This is irrelevant
in our case, since we will be interested in the limit $l \to \pm \infty$.
It remains, however, a very interesting problem to determine the spectral
flow of this theory in the interval $[-\half,\half]$.

\chapter{The spin chain system}

We come now to our main object. As was pointed out in [\Pol], in the strong
coupling constant limit the coordinate degrees of freedom of this system
decouple from the spin ones; this is because, for very large $l$, the
repulsion between particles becomes dominant and, unless a very high number
of quanta are excited, the coordinates of the particles assume their
classical equilibrium value. Thus, in this limit the coefficients of the
spin couplings become nondynamical constants and the spin degrees of freedom
decouple. Therefore, the above system becomes the tensor product
of a spinless Calogero system and of a spin chain system, with
spins lying at the equilibrium positions of the classical Calogero system.
Note that $\omega$ should also be scaled by a factor of $|l|$
in order to have a nontrivial limit. Dropping an overall scaling factor
of $|l|$, then, the hamiltonian of the spin chain system is
$$
H_s = -sgn(l) \sum_{i<j} {\sigij \over (x_i - x_j )^2} ,
\eqn\hams$$
where the $x_i$ minimize the Calogero potential, that is,
$$
\omega x_i - \sum_{i<j} {2 \over ( x_i - x_j )^3} = 0 ~.
\eqn\minim$$
The energy levels of the full system can be expressed as
$$
E_{s,p} = E_s + E_p ,
\eqn\EEE$$
where $E_s$ are the energy levels of the spin chain system and $E_p$ are
the levels of the Calogero system. Obviously, the states corresponding
to $E_p$ carry no spin, and therefore the full spin content of the level
$E_s$ coincides with the one of $E_{s,p}$. Since both $E_{s,p}$ and
$E_p$ have an equidistant spectrum, it already follows that the spectrum
of the spin chain $E_s$ is also equidistant.

The nice feature of the spectrum of the full system is that it is essentially
the same for all values $|l|>\half$, namely the spectrum of a free system,
depending nontrivially only on the sign of $l$.
Indeed, the spacing of the levels contains an overall factor $|l|$
(due to $\omega \to |l| \omega$), which we will drop as we did in \hams.
We will also take $\omega =1$ from now on. There is also an $l$-dependent
overall shift of the spectrum, due to the ground state energy,
which is irrelevant for the factorization \EEE. Thus,
the factorization \EEE\ holds even for {\it finite} values of $l$ (of course,
the wavefunctions in general will only factorize in the $|l| \to
\infty$ limit). Therefore, we reached the conclusion that the energy
eigenstates of the above spin chain system can be found by taking the states
of a free $N$-particle system with $n$ flavors and ``modding" them by the
states of a corresponding system with one flavor. Taking the particles
to be bosons (fermions) allows us to examine the spin system from the
ferromagnetic (antiferromagnetic) point of view, respectively. In what
follows we will implement this procedure to explicitly find the spin
system's spectrum and partition function.

Let $Z_{n,N}^B$ ($Z_{n,N}^F$) be the partition function of the bosonic
(fermionic) $N$-particle $n$-flavor problem for inverse temperature $\beta$,
and $Z_{n,N}^f$ ($Z_{n,N}^{af}$)
the partition function of the spin chain ferromagnetic (antiferromagnetic)
system. Then, \EEE\ implies that
$$
Z_{n,N}^B = Z_{n,N}^f Z_{1,N}^B
\eqn\ZZZ$$
and similarly for the $F,af$ case. For the bosonic case, dropping the
zero-point energy of the harmonic oscillator, the $N$-boson partition
function is given by
$$
Z_{1,N}^B = \sum_{k_i \leq k_{i+1}} \prod_{i=1}^N q^{k_i} ~,
\eqn\ZBB$$
where $q = e^{-\beta}$. By changing variables to $p_i = k_i - k_{i-1}$,
$p_1 = k_1$, we obtain
$$
Z_{1,N}^B = \sum_{p_i} \prod_{i=1}^N q^{(N-i+1) p_i}
= \prod_{k=1}^N {1 \over 1- q^k} ~.
\eqn\ZB$$
To find $Z_{n,N}^B$, we remark that the grand canonical partition function
of the $n$-flavor bosonic system is the product of $n$ one-flavor ones, thus
$$
{\cal Z}_n^B = \sum_{N=0}^\infty Z_{n,N}^B y^N = ({\cal Z}_1^B )^n = \left(
\sum_{N=0}^\infty Z_{1,N}^B y^N \right)^n ~,
\eqn\GP$$
where $y=e^{-\mu}$ with $\mu$ the chemical potential. Therefore
$$
Z_{n,N}^B = \sum_{\sum_i k_i =N} \prod_{i=1}^n Z_{1,k_i}^B ~.
\eqn\Zpr$$
Combining \ZZZ, \ZB\ and \Zpr\ we finally obtain
$$
Z_{n,N}^f = \sum_{\sum_i k_i =N}
{\prod_{k=1}^N (1- q^k ) \over \prod_{i=1}^n \prod_{r=1}^{k_i} (1- q^r )} .
\eqn\Zf$$

{}For the fermionic case, the single-flavor partition function is identical
to the bosonic one (this is a particular case of bosonization in one
dimension, since the oscillator spectrum is ``relativistic"), the only
difference being the ground state energy. Therefore,
$$
Z_{1,N}^F = q^{{N(N-1) \over 2}} \prod_{k=1}^N {1 \over 1- q^k } ~.
\eqn\ZF$$
The rest of the argument is the same as in the bosonic case. (Note that
the inclusion of the ground state energy in \ZF\ is necessary to get the
correct $n$-flavor partition function in the fermionic version of \Zpr.)
We obtain
$$
Z_{n,N}^{af} = q^{-E_{n,N}^{af}} \sum_{\sum_i k_i =N} \prod_{k=1}^N (1- q^k )
\prod_{i=1}^n {q^{\half (k_i -{N\over n})^2} \over
\prod_{r=1}^{k_i} (1- q^r )} ~.
\eqn\Zaf$$
In the above, we had to subtract explicitly the zero-point energy
of the partition function, so as to bring the ground state
to zero energy. This was necessary since the fermionic systems used to
derive \Zaf\ had a nonzero ground state energy. The constant $E_{n,N}^{af}$
is calculated to be
$$
E_{n,N}^{af} = {m(n-m) \over 2n}~~,~~~~m = N\, {\rm mod}\, n .
\eqn\Eaf$$

A number of nontrivial checks can be performed on the above partition
functions. Firstly, since they are partition functions of a finite system
with equidistant levels, they should be a polynomial in $q$.
This may not be obvious from the expressions \Zf\ and \Zaf,
but it will be proved in a short while. Next,
the value of either $Z_{n,N}^f$ or $Z_{n,N}^{af}$ for $q=1$ should
reproduce the number of states of the system. Indeed, in
that limit, \Zf\ and \Zaf\ simply become the polynomial expansion of
$$
n^N = (1+1+ \cdots 1)^N = \sum_{\sum_i k_i =N}
{N! \over \prod_{i=1}^n k_i !} ~.
\eqn\Nst$$
Finally, the two partition functions really describe the same problem,
with the spectrum reversed. This means that
$$
Z_{n,N}^{af} (q) = q^{E_{max}} Z_{n,N}^f ( q^{-1} )
\eqn\ZqZ$$
where $E_{max}$ is the energy of the highest excited state in either
case. By isolating the highest power of $q$ in \Zf, we find
$$
E_{max} = {n-1 \over 2n} N^2 - {m(n-m) \over 2n} .
\eqn\Emax$$
It is easy to check that \ZqZ\ indeed holds for the expressions \Zf\
and \Zaf.

\chapter{The $SU(2)$ case}

For the special case of $SU(2)$, the partition functions take the form
$$
Z_{2,N}^f = \sum_{k=0}^N \prod_{r=1}^k {1- q^{N-r+1} \over 1- q^r} ,
\eqn\Zff$$
$$
Z_{2,N}^{af} = q^{-{m\over 4}}  \sum_{k=0}^N
q^{(k-{N\over 2})^2} \prod_{r=1}^k {1- q^{N-r+1} \over 1- q^r} .
\eqn\Zaff$$
Formula \Zaff\ reproduces the partition function proposed in [\Fra]
on the basis of numerical investigations (modulo a ground-state energy
correction for odd $N$).

Some properties of the $SU(2)$ spectrum are
easy to infer directly from the corresponding particle system.
{}From the ferromagnetic end, it is clear that the first irrep of spin
${N\over 2} - k$ will appear at energy $k$, coming from the state where
$k$ bosons have been excited to the first excited state, which carries
the representation $({N-k \over 2})\otimes ({k \over 2})$ (numbers in
parentheses correspond to spin). Similarly, from
the antiferromagnetic end, the first irrep of spin ${m\over 2}+k$ will
appear at energy $k(k+m)$, corresponding to $k$ fermions excited to
successive levels above the Fermi level and leaving $k$ ``holes" behind,
which carries the representation $(\half)\otimes \cdots (\half)$ ($2k$ times).
The creation operator picture is also useful; clearly the set of all
polynomials in $A_k$, $A_k^a$, modded by the set of all polynomials in
$A_k$ (the single-flavor case), leaves the set of all polynomials in
$A_k^a$. Each such operator, acting on a state of total spin $s$ will
produce a state of spin $s-1$ [\MiPo], as long as $s\geq 1$. The states of
the first ${N\over 2}$ energy levels in the ferromagnetic case can be
generated this way. Counting all possible monomials of degree $k$, we find
that the subspace of the Hilbert space with energy $k$ decomposes as
$$
{\cal H}_k = \sum_{i=1}^k \oplus \Bigl({N\over 2} -i\Bigr)_{p(k,i)} ~,
\eqn\Hk$$
where $p(k,i)$ denotes the number of partitions of $k$ into $i$ positive
integers and counts the multiplicity of each spin.

We emphasize here that the partition functions \Zff\ and \Zaff\ reproduce the
{\it full representation content} of the energy levels. To see this, remember
that the sum over $k$ in the above partition functions came from the sum
over $k$ in \Zpr\ (for $N=2$). Therefore, each term represents a state
with $k$ particles of the one flavor and $N-k$ particles of the other.
Since ``flavors" correspond to the different spin states of the particles,
each term $k$ corresponds to states with total 3-component of spin
$S^3 = k \half + (N-k) (-\half) = k -{N \over 2}$.
Thus, the above sum groups the energy
levels in terms of the 3-component of their total spin $S^3$. Once we have
all the states in each energy level and their spins, there is always a
unique way to group them into representations of the total spin.

Having realized that, it is trivial to write the partition function of the
system in the presence of a magnetic field $B$ in the 3-direction, as
$$
Z_{n,N}^f (q,w) = \sum_{k=0}^N w^{(k-{N\over 2})} \prod_{r=1}^k
{1- q^{N-r+1} \over 1- q^r} ,~~~~w=e^{-\beta B}=q^B ~,
\eqn\ZffB$$
and similarly for the antiferromagnetic case. An interesting property of the
system stemming from the above is the following: if a ``global" interaction
between the spins is included, of the form $-(S^3)^2$, then the
antiferromagnetic system is transformed into the ferromagnetic one.
This is very peculiar: the presence of such an interaction breaks the
global $SU(2)$ invariance. Therefore we do not expect the states to fall
into $SU(2)$ multiplets, while they obviously do in this particular case
since the spectrum is identical to the $SU(2)$-invariant ferromagnet.
Obviously the multiplets obtained this way will not be representations
of the global $SU(2)$, but there should be some nontrivially-defined set
of $SU(2)$ generators in this case which commute with the hamiltonian.

A useful observation about the partition function \ZffB\ is that has the
form of a $q$-binomial. We remind the reader that the $q$-deformation of
a number is, in one definition [\Ext],
$$
[x]_q = {q^x -1\over q-1}
\eqn\q$$
and the $q$-factorial
$$
[k!]_q = \prod_{r=1}^k [r]_q ~.
\eqn\qfac$$
{}From this, we define the $q$-($m$-choose-$n$) symbol
$$
\left[ \matrix{m \cr n \cr}\right]_q = {[m!]_q \over [n!]_q [(m-n)!]_q}
\eqn\qchoice$$
and the $q$-binomial as
$$
[1+w]_q^N = \sum_{k=1}^N \left[\matrix{N\cr k\cr}\right]_q w^k .
\eqn\qbin$$
It is easy to see that, up to an overall factor $w^{-{N\over 2}}$, the
ferromagnetic $Z_{2,N}^f (q,w)$ is exactly the $q$-binomial of the $N$-th
degree for $w$. This is useful in finding a factorized form for $Z_{2,N}^f$.
Indeed, for an alternative symmetric definition of the $q$-deformation
$$
[[x]]_q = {q^{{x \over 2}}- q^{-{x \over 2}} \over q^\half - q^{-\half}}
= q^{1-x \over 2} [x]_q ,
\eqn\qalt$$
it is known that the corresponding $q$-binomial assumes the form
$$
[[1+ q^{{N-1 \over 2}} w]]_q^N = \prod_{k=0}^{N-1} (1+q^k w) .
\eqn\qbinmod$$
The above formula admits the interpretation of the grand canonical
partition function of a fermionic system with levels $E=0,1, \dots N-1$
and chemical potential $e^{-\mu}=w$. Using \qalt, on the other hand, we can
see that the ferromagnetic partition function differs from the above
$q$-binomial by an extra factor $q^{k(k-1)\over 2}$ in each term. From
\qbinmod, then, we are led to an effective description of the system as
a fermionic system with $N$ levels and hamiltonian
$$
H_{eff} = \sum_{k=0}^N (k+B) n_k - \sum_{k<l} n_k n_l
\eqn\Heff$$
where the occupation numbers $n_k$ take the values $0,1$. The corresponding
antiferromagnetic effective hamiltonian can be found by taking $H_{eff}
\to -H_{eff}$ and adding $E_{max} = {N^2 \over 4} -{m\over 4}$.
This reproduces, for $B=0$, the model derived in [\Fra] on the basis of
numerical results. Other useful properties of the partition function can
be derived by using $q$-function results. For instance, by acting with a
$q$-derivative on $Z_{2,N}^f (B)$ we find the recursion relation
$$
Z_{2,N}^f (B+1) = Z_{2,N}^f (B) - [N]_q \, q^{B\over 2} \, (1-q) \,
Z_{2,N-1}^f (B) ~.
\eqn\rec$$
A corresponding relation for the antiferromagnetic case is easily obtained.

\chapter{The $SU(n)$ case}

The general $SU(n)$ case can be examined in an analogous way. Firstly,
the partition function \Zf\ (or \Zaf) also fully determines the $SU(n)$
representation content of the spectrum. The ``flavor" of the particles in
$Z_{n,N}^B$ corresponds to the $n$ different components of the $SU(n)$
spin of each particle, and therefore the set of $n-1$ independent quantities
$k_i -{N\over n}$ determines the values of the $n-1$ Cartan generators
of $SU(n)$, in the nonorthogonal fundamental parametrization
$$
(H^i)_{jk} = \delta_{ij} \delta_{ik} - {1 \over n}
\eqn\Cartan$$
(the above are just the $U(n)$ Cartan generators minus $Q \over n$, with
$Q$ the $U(1)$ charge).
These are enough to fully determine the $SU(n)$ state. Once all
these states are known for each energy level, they can be assembled into
irreps of $SU(n)$. We can also introduce an $SU(n)$ magnetic field
interaction in the problem, as
$$
H_{mag} = \sum_{a=1}^{N^2 -1} S^a B^a .
\eqn\Hmag$$
With a global $SU(n)$ transformation we can always rotate $B^a$ in the
Cartan subspace, and the interaction takes the symmetric form
$$
H_{mag} = \sum_{i=1}^N k_i B_i ~~~~{\rm with}~~~\sum_i B_i =0 .
\eqn\Hkmag$$
So the full partition function takes the form
$$
Z_{n,N}^f (q,\{ w_i \}) = \sum_{\sum_i k_i =N} \left[ \matrix{ N \cr
\{ k_i \} \cr} \right]_q \prod_{i=1}^n w_i^{k_i}
\eqn\Zfw$$
where $w_i = e^{-\beta B_i}$ and we defined the generalized $q$-choose
symbol
$$
\left[ \matrix{ N \cr \{ k_i \} \cr} \right]_q = {[N!]_q \over
\prod_i [k_i !]_q} .
\eqn\qqchoose$$
Similarly, for the antiferromagnetic case we obtain
$$
Z_{n,N}^{af} (q,\{ w_i \}) = q^{-E_{n,N}^{af}} \sum_{\sum_i k_i =N}
\left[ \matrix{ N \cr \{ k_i \} \cr} \right]_q \prod_{i=1}^n
q^{(k_i -{N\over n})^2} w_i^{k_i} .
\eqn\Zafw$$
Again we observe that an extra interaction of the form $\sum_i (H^i)^2$
turns the system from ferromagnetic to antiferromagnetic.

{}From the above, it follows that the $SU(n)$ partition function is the
$q$-polynomial of the $N$-th degree in the variables $w_i$. A systematic
way of writing this polynomial, other than \Zfw, is the following:
consider the variables $X_i$, $i=1,\cdots n$, obeying the
$q$-commutation relations
$$
X_i X_j = q X_j X_i ~~~~{\rm for}~~~i>j .
\eqn\AA$$
Then the {\it ordinary} polynomial expansion of $(X_1 + \cdots X_n )^N$
automatically reproduces the $q$-polynomial, once the variables in each
monomial are ordered in ascending index order. This proves, in particular,
that the partition function is a polynomial in $q$ since the above will
manifestly produce polynomial coefficients.

The previous construction allows us to find
a model hamiltonian for this system. Consider $N$ levels, all at zero
energy, ordered from 1 to $N$. In each level place exactly one particle
of flavor $i=1,\cdots n$, and for every ``reversal of order" in the
configuration increase the energy by one. That is, for each pair of particles
such that the one of a lower flavor is in a higher level than the other, the
energy increases by one. The model hamiltonian in terms of occupancies
$n_{i,k}$, $i$ labeling the flavor and $k$ the level, takes the form
$$
H_{eff}^f = \sum_{i,k} B_i n_{i,k} + \sum_{k<l,\, i>j} n_{i,k} n_{j,l} ~,
{}~~~n_{i,k}=0,1~,~~~\sum_i n_{i,k} =1.
\eqn\Hefff$$
Alternatively, we can describe the system in terms of parafermions of
order $n$, the flavor index of the previous model becoming the occupancy.
In this language, the hamiltonian becomes
$$
H_{eff} = \sum_k B_{1+n_k} +\sum_{k<l} \theta ( n_k - n_l ) , ~~~
n_k = 0,1,\cdots n-1 ,
\eqn\Hpar$$
with the step function $\theta$ defined
$$
\theta ( x ) = \left\{ \matrix{1~~{\rm for}~x>0,\cr
0~~{\rm for}~x\leq 0 . \cr}\right.
\eqn\theta$$
One can see that the case $n=2$ is, up to a constant, equivalent to the model
given earlier in \Heff\ with $B_1 = -B_2 = {B \over 2}$.
Therefore, the states of the $SU(n)$ problem can be described in
terms of $N$-digit numbers in the $n$-basis (``modules") $n_1 \cdots n_N$,
which determine the energy as well as the $SU(n)$ quantum numbers
of the state. In terms of these modules and the hamiltonian \Hpar, it
follows that the ground state consists of all totally ordered configurations
$0 \dots 1 \dots (n-1)$, with an otherwise arbitrary multiplicity of each
occupancy. There are obviously ($N+n-1$-choose-$n-1$) such configurations,
reproducing the degeneracy of the ferromagnetic ground state. The highest
excited state is the one with the ``maximal reversal of order," that is,
the state $(n-1) \dots 1 \dots 0$, with a multiplicity of each occupancy
as close to $N\over n$ as possible. There are ($n$-choose-$m$) such
modules, and this reproduces the degeneracy of the antiferromagnetic
ground state.

\chapter{Epilogue}

In conclusion, we see that the above system exhibits a simple but very
rich structure. Unlike the standard Haldane-Shastry model, its spectrum is
equally-spaced and therefore simpler. The highly degenerate structure of the
levels, on the other hand (``supermultiplets") is a feature common to the
two models, although they differ in the actual representation content of
the levels. We also remark that the present system does not have a (lattice)
translation invariance. It is therefore remarkable that, in spite of this
fact, it is amenable to a complete solution, for arbitrary $n$ and $N$.
There are still, however, many
interesting issues to be addressed. Among them are a systematic,
group-theoretic procedure of finding the level structure in the
antiferromagnetic regime, its relation to conformal field theory,
the statistical mechanics of the $SU(n)$-magnetic field model,
and the identification and study of spinon excitations.

\ack{I am thankful to A.M.~Perelomov for reviving my interest in this
problem, to C.~Zachos for a crash-tutorial on $q$-functions and their
properties, and to J.~Minahan for a critical reading of the manuscript.}
\refout
\end